\documentstyle[prl,aps,floats]{revtex}
                                      
\begin{document}

\title{\rightline{Fermilab-Pub-01/258-E}\vspace{0.5cm}Hadronic Production of $\Lambda_c$ from 600 GeV/c $\pi^-$, $\Sigma^-$ and $p$ beams}

\author{
F.G.~Garcia$^{19,5}$,
G.~Alkhazov$^{11}$,
A.G.~Atamantchouk$^{11}$$^{,\ast}$,
M.Y.~Balatz$^{8}$$^{,\ast}$,
N.F.~Bondar$^{11}$,
P.S.~Cooper$^{5}$,
L.J.~Dauwe$^{17}$,
G.V.~Davidenko$^{8}$,
U.~Dersch$^{9}$$^{,\dag}$,
A.G.~Dolgolenko$^{8}$
% ...
\\                                                                            
\vskip 0.50cm                                                                 
\centerline{(SELEX Collaboration)}                                             
\vskip 0.50cm                                                                 
}                                                                             
\address{
% ...
$^5$Fermilab, Batavia, IL 60510, U.S.A.\\
% ...
$^8$Institute of Theoretical and Experimental Physics, Moscow, Russia\\
$^9$Max-Planck-Institut f\"ur Kernphysik, 69117 Heidelberg, Germany\\
% ...
$^{11}$Petersburg Nuclear Physics Institute, St. Petersburg, Russia\\
% ...
$^{17}$University of Michigan-Flint, Flint, MI 48502, U.S.A.\\
% ...
$^{19}$University of S\~ao Paulo, S\~ao Paulo, Brazil\\
% ...
}
\date{\today}
\maketitle
%
%
%
%
\begin{abstract}

We present data from Fermilab experiment E781 (SELEX) on the hadroproduction asymmetry for $\overline{\Lambda}_c^{~-}$ compared to $\Lambda_c^+$ as a function of $x_F$ and $p_t^2$ distributions for $\Lambda_c^+$. These data were measured in the same apparatus using incident $\pi^-$, $\Sigma^-$ beams at 600 GeV/$c$ and proton beam at 540 GeV/$c$. The asymmetry is studied as a function of $x_F$. In the forward hemisphere with $x_F \ge 0.2$ both baryon beams exhibit very strong preference for producing charm baryons rather than charm antibaryons, while the pion beam asymmetry is much smaller. In this energy regime the results show that beam fragments play a major role in the kinematics of $\Lambda_c$ formation, as suggested by the leading quark picture.

\end{abstract}

\twocolumn
\input{psfig}
%
%
%
%
\section{Introduction}

The conventional leading order (LO) and next to leading order (NLO) charm-anticharm production diagrams at the quark level show little or no asymmetry in the $x_F$ or $p_t$ behaviour of the quark and antiquark, as viewed in the laboratory frame. Previous studies, primarily using pion beams at a variety of energies up to 500 GeV/$c$, have shown that in many cases charm mesons that share a quark in common with the beam hadron, so-called {\sl leading charm}, are produced more copiously in the forward hemisphere than those states which do not have a shared quark, i.e., {\sl non-leading charm}~\cite{E791}. Recently, this has been extended to charm baryon production by a $\Sigma^-$ beam, where asymmetry for $\Lambda_c^+$ compared to $\overline{\Lambda}_c^{~-}$ as a function of $x_F$ was reported to be large~\cite{WA89}. There are limited-statistics proton data available for central production, where only a 90$\%$ confidence level lower limit is reported~\cite{E769}.

Because the quark-level processes are thought to be symmetric, the asymmetries are viewed as features of the hadronization process~\cite{lund} or as a manifestation of an intrinsic charm content of the beam hadron~\cite{intrinsic}. In this experiment the $\Lambda_c^+$ shares a quark in common with all three beam hadrons, while the antibaryon has an antiquark in common with the $\pi^-$ beam only. We shall compare the hadroproduction characteristics for these charm baryon and antibaryon states to illuminate the role of the beam fragments in hadronization induced asymmetries.
%
%
%
%
\section{Apparatus Description}

The SELEX spectrometer is a three-stage magnetic spectrometer designed to cover charm production in the forward hemisphere. The beam hadron was identified by a transition radiation detector as pion or not-pion and tracked in 8 beam silicon strip detectors before interacting in a set of 5 target foils, 2 copper and 3 carbon totalling 5$\%$ of a proton interaction length. Interaction products were tracked in a set of 20 vertex silicon detectors arranged in 4 sets of planes rotated by $45^0$. Single-track efficiencies for each system exceeded 98$\%$. Data were taken with negative beam at 615 GeV/$c$ (half $\pi$, half $\Sigma$) totalling 12.3 $\times~10^9$ inelastic interactions and positive beam at 540 GeV/$c$ (92$\%$ protons) totalling 2.8 $\times~10^9$ inelastic interactions. The layout of the spectrometer can be found elsewhere~\cite{layout}.\\ \indent

Two Proportional Wire Chamber (PWC) magnetic spectrometer systems analyzed tracks with momenta above 3 GeV/$c$ (M1 spectrometer, 12 planes) or 15 GeV/$c$ (M2 spectrometer, 14 planes). The beam region of the PWCs was covered with double-sided silicon detectors to give high-resolution momentum measurement up to full beam energy. The combination of excellent vertex-region tracking resolution (typically 65 $\mu$rad at 150 GeV/$c$) and excellent momentum resolution ($\sigma_p/p \approx 0.5\%$ for a typical 100 GeV/$c$ track) gave a ${\Lambda_c}^+$ mass resolution ($\sigma$) of about 9 MeV/$c^2$ at all $x_F$. Tracks above 22 GeV/$c$ traversed a Ring-Imaging \v{C}herenkov (RICH) and were identified as $\pi$, K, or p~\cite{rich-nim}. The RICH distinguished $K^{\pm}$ from $\pi^{\pm}$ up to 165 GeV/$c$. The proton identification efficiency was greater than 95 $\%$ above proton threshold ($\approx$ 90 GeV/$c$) and greater than 90 $\%$ above kaon threshold ($\approx$ 45 GeV/$c$).

An important innovation in SELEX was the use of an online topological trigger to identify charm. The hardware trigger was loose, requiring $\ge 4$ charged hadrons in the forward 150 mrad cone and $\ge 2$ hits from positive track candidates in a hodoscope after M2. About 1/3 of all inelastic interactions satisfied this trigger. The software trigger made a full vertex reconstruction of the beam track and all tracks in the M2 spectrometer (high momentum tracks) to test that hypothesis that they all came from a single primary vertex. A software-adjustable $\chi^2$ cut of 8.5 selected candidate events that were inconsistent with a single primary vertex, i.e., events that {\sl might} have a downstream decay. This gave a rejection factor of about 8 in the data volume. Studies with the software trigger turned off show that this costs about a factor of 2 in charm efficiency, for a net filter enhancement of 4.  
%
%
%
%
\section{Data Analysis}

A general data reduction pass found reconstructable charged tracks and identified the interacting beam track. Primary and secondary vertices were found by geometric reconstruction without regard to particle identification. The high resolution of the beam and vertex systems, along with the thin production targets, gave typical primary vertex resolution of 270 $\mu$m and secondary vertex resolution for these $\Lambda_c^+$ events of 560 $\mu$m for a mean $\Lambda_c^+$ momentum of 230 GeV/$c$.

All the $p K \pi$ candidates used in this analysis were selected by the following requirements:
 (1) primary and secondary vertex fits each had $\chi^2$/dof $<$ 5;
 (2) the primary-secondary vertex separation significance (L/$\sigma$) had to be greater than 8, where $\sigma$ is the quadrature sum of the primary and secondary vertex errors;
 (3) the total momentum vector of the decay tracks had to point back to the primary vertex within errors; 
 (4) ($\overline{\Lambda}_c^{~-}$) $\Lambda_c^+$ events were required to have positive RICH identification for the (anti)proton and the ($K^+$)$K^-$. The pion could be any track;
 (5) secondary vertices which occured inside any target were removed;
 (6) each secondary track was extrapolated back to the primary vertex to determine its transverse miss distance. The second-largest transverse miss distance had to exceed 20 $\mu$m.

\begin{figure}[h]
\centerline{\psfig{figure=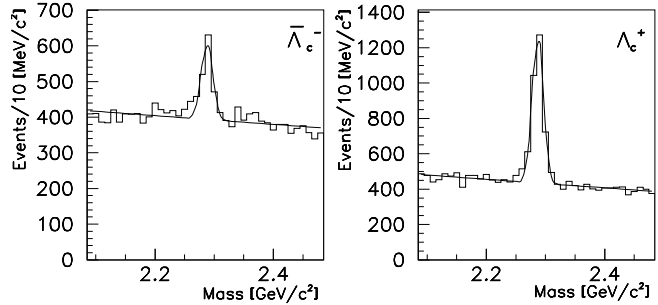,width=90mm}}
\caption{Mass distributions for $\overline{\Lambda}_c^{~-}$ (left) and $\Lambda_c^+$ (right) where signals for all beams are combined. Gaussian fits are shown for illustration only.}
\label{fig:mass}
\end{figure}

Events were selected by the above cuts and sorted by the primary beam tag: $\Sigma^-$ or $\pi^-$ from the negative beam and protons from the positive beam. Figure~\ref{fig:mass} shows the full SELEX statistics for the $\overline{\Lambda_c}^{~-} \rightarrow \overline{p} K^+ \pi^-$ and  $\Lambda_c^+ \rightarrow p K^- \pi^+$ mass plots where signals for all beams are combined.
\indent Gaussian fits are shown for illustration only. The background shapes are smooth on the low mass and high mass side of the signal regions. The signal yield for bayons and antibaryons are determined by a sideband subtraction method which assumes a linear mass resolution dependence for the momentum spectrum. The number of background events estimated outside the signal window is subtracted from the number of events in the signal window. The signal window corresponds to a 3.33-$\sigma$ mass width and the background is 10-$\sigma$ centered at 2.285 GeV/$c^2$. We find  $1979 \pm 71$ signal events for $\Lambda_c^+$ and $520 \pm 60$ signal events for $\overline{\Lambda}_c^{~-}$ 
%
%
%
%
\section{Asymmetry Analysis and $x_F$ Distributions}

The data yields were corrected in each $x_F$ bin by the total acceptance (geometrical acceptance and reconstruction efficiency). The simulation used Monte Carlo generated $\Lambda_c^+$ (or c.c.) events with phase space decay to the $p K \pi$ final state. The simulated trajectories were converted to hits in all tracking detectors, including multiple Coulomb scattering but not secondary interaction losses. The hits were embedded in the hit banks of real data and all tracks were reconstructed. This overestimates slightly the tracking inefficiency, but this effect is small. The RICH simulation has been tested against a set of $\Lambda^0 \rightarrow p + \pi^-$ events taken without a RICH cut. Agreement is excellent; the proton identification efficiency averaged over the momentum spectrum is 95$\%$.

\begin{figure}[h]
\centerline{\psfig{figure=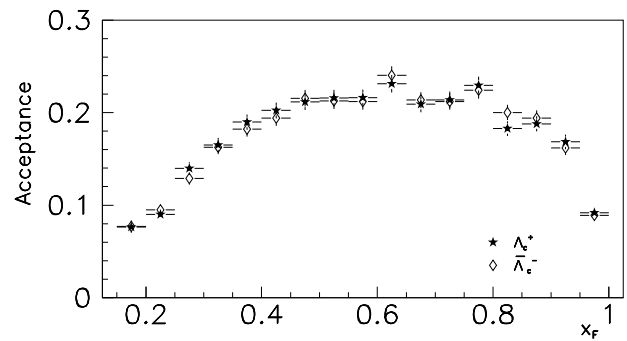,width=90mm}}
\caption{Acceptance of $\Lambda_c$ reconstruction.}
\label{fig:accep}
\end{figure}

\begin{figure}[h]
\centerline{\psfig{figure=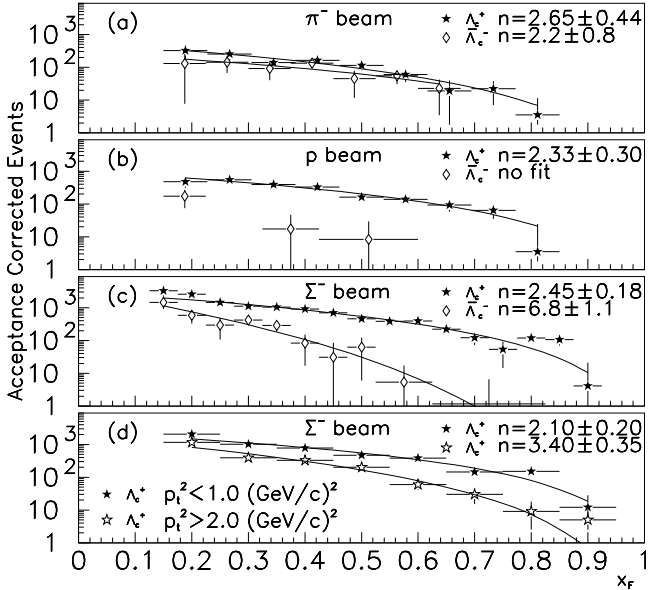,width=90mm}}
\caption{$\Lambda_c$ $x_F$ distribution of corrected number of events for $\pi^-$ (a), proton (b) and $\Sigma^-$ (c) beams. (d) shows the $\Lambda_c^+$ $x_F$ distribution for $\Sigma^-$ production at low $p_t^2$ and high $p_t^2$.}
\label{fig:xf}
\end{figure}

\begin{figure}[h]
\centerline{\psfig{figure=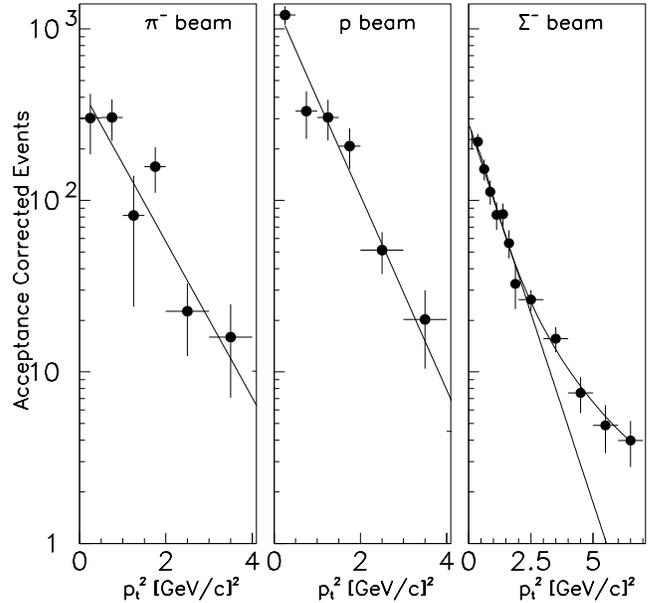,width=90mm}}
\caption{$\Lambda_c$ $p_t^2$ distribution of corrected number of events for $\pi^-$ (left), proton (center) and $\Sigma^-$ (right) beams.}
\label{fig:pts}
\end{figure}

The simulated events were analysed with the same tracking and reconstruction procedure used in the data. The whole set of cuts used to extract the signal was applied here as well. The acceptance is the ratio between the number of events reconstructed by the number of embedded events generated. SELEX acceptance does not depend on the transverse momentum of the charm particle up to $p_t^2=7$ (GeV/$c$)$^2$. Therefore we consider only the acceptance as a function of $x_F$. An important issue for this analysis is the relative efficiency for the baryon and antibaryon decays. As seen in figure~\ref{fig:accep}, there is no discernable difference in the acceptances. The $x_F$ distribution of the acceptance-corrected number of events from each beam is shown in figure~\ref{fig:xf}. Because the beam flux is the same for baryon and antibaryon, these figures compare the relative production cross sections for these states. The curves shown are fits of the standard parametrization $(1-x_F)^n$ to the data. The results are summarized in table~\ref{tab xF}.
\begin{table}
\begin{tabular}{c c c c c}
  Beam    & \multicolumn{2}{c}{$\Lambda_c^+$} & \multicolumn{2}{c}{$\overline{\Lambda}c^{~-}$}\\
 particle & $x_F$       & $n-value$       & $x_F$      & $n-value$ \\ \hline
$\pi^-$        & 0.15 - 0.85 & $2.65 \pm 0.44$ & 0.15 - 0.675 & $2.2 \pm 0.8$ \\
$\Sigma^-$     & 0.125 - 0.925 & $2.45 \pm 0.18$ & 0.125 - 0.825 & $6.8 \pm 1.1$ \\
$p$            & 0.15 - 0.85 & $2.33 \pm 0.30$ & 0.15 - 0.60 & no fit 
\end{tabular}
\caption{Summary of measured n-value from the fit $(1-x_F)^n$ for $\Lambda_c^+$ and $\overline{\Lambda}_c^{~-}$. The errors are statistical only.}
\label{tab xF}
\end{table}
The relative sizes of the differential cross sections for baryon and anti-baryon production are clearly different for pion and baryon beams. The pion beam valence quark content is baryon/antibaryon symmetric. Relative cross sections for $\Lambda_c^+$ and $\overline{\Lambda}_c^{~-}$ production from pions are comparable in shape and magnitude over the entire $x_F$ range.  Such behaviour is consistent with the leading particle picture, since both charm states share a valence quark in common with the beam.  For $\Sigma^-$ and proton beams  $\Lambda_c^+$ production is much stronger than $\overline{\Lambda}_c^{~-}$ production, especially at high $x_F$. In this case there are no valence antiquarks, and antibaryon production appears to be strongly disfavored. The four distributions that show leading behaviour are consistent with a common $(1-x_F)^n$ dependence having n $\sim$ 2.5, much harder than the n for leading meson production from pion beams~\cite{ptref,meson-n}. Note that the high-statistics $\Sigma^-$ distribution shows some interesting structures. 
Clearly $\Lambda_c^+$ production is favored over $\overline{\Lambda}_c^{~-}$, and the enhancement increases with $x_F$.
%
%
%
%
The $p_t^2$ spectrum for $\Lambda_c^+$ production by the three different beam hadrons is shown in Fig.~\ref{fig:pts}.  For $p_t^2 \le$ 2 (GeV/$c$)$^2$ all three beams are consistent with a smooth exponential spectrum $e^{-bp_t^2}$ with the same parameter b $\approx$ 1.1 (GeV/$c$)$^{-2}$. At larger $p_t^2$ values, seen only in the high-statistics $\Sigma^-$ data, the exponential gives way to a power-law behavior as expected in QCD and as seen in $\rm{D}^{\pm,0}$ production~\cite{ptref}. There is no indication in the large-$x_F$ events of an enhancement at small $p_t^2$ as suggested by the intrinsic charm picture. However, when we look at the $x_F$ behaviour of events after the slope change at $p_t^2 \ge$ 2 (GeV/$c$)$^2$, we see a significant difference in the $x_F$ behaviour. The events with large $p_t^2$ have a softer $x_F$ spectrum than those at small $p_t^2$, as seen in figure~\ref{fig:xf} (d).  
The hadroproduction asymmetry A is defined as:
\begin{eqnarray}
A \equiv {\sigma_{c} - \sigma_{\overline{c}} \over 
  \sigma_{c} + \sigma_{\overline{c}}}
\end{eqnarray}

\noindent where $\sigma_{c}$ ($\sigma_{\overline{c}}$) is the production cross-section for the charm particle (anti-charm particle) in study.

We calculated the values of the asymmetry and the errors using the maximum likelihood method with no limit at $A = \pm 1$. Events in each $x_F$ bin follow the Poisson distribution. Therefore, the likelihood function is a product between two Poisson distributions, one for the background region and the other for the signal region.

\begin{figure}[h]
\centerline{\psfig{figure=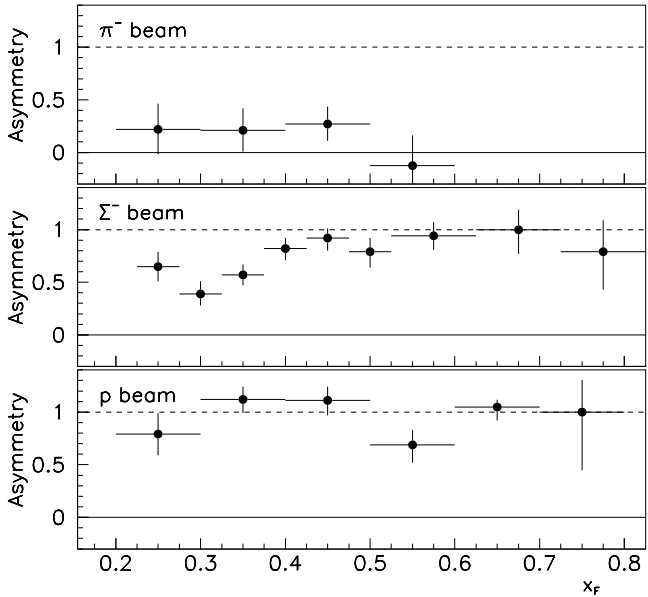,width=90mm}}
\caption{Asymmetry for $\Lambda_c$ production by $\pi^-$ (top), $\Sigma^-$ (center) and proton (bottom) beam.}
\label{fig:asy}
\end{figure}

The results are presented for the three beam hadrons in figure~\ref{fig:asy}.
As is already clear from the $x_F$ distributions, $\Lambda_c$ production in $\Sigma^-$ and proton beams shows a strong asymmetry favoring baryons over antibaryons at all $x_F$. The proton asymmetry rises sharply toward unity already at small $x_F$, reflective of the paucity of  $\overline{\Lambda}_c^{~-}$ production by protons. The $\Sigma^-$ asymmetry has a smoother approach to the limit at $+1$. In the case of the $\pi^-$ beam we see only a slightly positive and constant asymmetry throughout the $x_F$ range. Our result agrees well in the overlap region with recent E791 data~\cite{proc}.  

\section{Systematic Errors}

Systematic errors in  kinematic analysis may arise from several sources.  We list here the ones that have been considered.  In all cases systematic effects are found to be small compared to statistical errors.

\begin{itemize}
\item $x_F$ binning effects\\
The $x_F$ distribution is relatively smooth, but there are structures in the high statistics data. We have shifted the bin limits for all $x_F$ distributions by 0.025 (half a bin) and refit the data. No effect in the character of the distributions is observed. For example, all the structures seen in Fig.3(c) are also present in the shifted distribution. Changes in the n-value are consistent with purely-statistical effects. Therefore, we quote a conservative $\Delta$n for the fit of 0.1, even though no systematic effect has been seen.\\

\item Mass window size\\
The signal window in the mass distribution is taken as $\pm$30 MeV/$c^2$
around 2.285 GeV/$c^2$. We have changed the window up to $\pm$35 MeV/$c^2$ and down to $\pm$25 MeV/$c^2$. The systematic error for changing the mass window compared to statistical error is negligible. 

\item Acceptance corretion \\
The acceptance shown in Fig. 2 is computed using the method of adding hits from simulated charm decay tracks into the data banks from an observed event. The modified event is reanalysed to see if the charm state is reconstructed. There could be a shift in the corrected $x_F$ distribution if the acceptance calculation does not represent truly the $x_F$ variation of the apparatus efficiency. In E781 the dominant efficiency variation at low $x_F$ is purely geometric: both proton and kaon pass through the fiducial volume of the RICH counter. A second geometric effect is the related one: the charm particle live long enough to pass the vertex significance cut. The tracking characteristics of the spectrometer match those predicted from the simulation very well over all momenta studied here. The $K_s$ mass is the same for all observed momenta, both in data and simulation. We believe that there is no systematic bias from the simulation and assign no error.

\item Trigger Bias in Asymmetry\\
The $x_F$-dependent effects tend to cancel in the asymmetry measurement.  However, a new systematic effect must be considered: Trigger bias. The E781 trigger required counts on the postive-bend side of a trigger hodoscope after the second bending magnet. This {\it might} favor $\Lambda_c^+$ production over $\overline{\Lambda_c^+}$. We looked at the asymmetry of $K_s$ production, for times when the $\pi^+$ is found in the high-momentum spectrometer compare to the converse. We see a maximum 3\% asymmetry. We also look at the charm baryon data itself. If we $\underline{\rm{exclude}}$ the charm tracks, then in 97\% of the triggers there are still enough hits in the hodoscope to trigger the event without the charm decay tracks. We conservatively assign a 3\% systematic error to the asymmetry from trigger bias.

\end{itemize}

\section{Summary}

The SELEX experiment reconstructed about 2500 $\Lambda_c \rightarrow p K \pi$ from 600 GeV/$c$ $\Sigma^-$ and $\pi^-$ beams and a 540 GeV/$c$ proton beam. We observed that the $x_F$-dependence of $\Lambda_c^+$ production is similar for all three beams, much stiffer than for meson production. Both baryon beams show a strong enhancement of the production of charm baryons over antibaryons, while the two are produced comparably from a pion beam. These results are in general agreement with color-drag models of hadronization. The differences in behaviour for protons and $\Sigma^-$ illustrate significant non-perturbative effects in production dynamics at these energies. 
%
%
%
%

The authors are indebted to the staff of Fermi National Accelerator 
Laboratory and for invaluable technical support from the staffs of collaborating institutions. This project was supported in part by Bundesministerium f\"ur Bildung, Wissenschaft, Forschung und Technologie, Consejo Nacional de 
Ciencia y Tecnolog\'{\i}a {\nobreak (CONACyT)}, Conselho Nacional de Desenvolvimento Cient\'{\i}fico e Tecnol\'ogico, Fondo de Apoyo a la Investigaci\'on (UASLP), Funda\c{c}\~ao de Amparo \`a Pesquisa do Estado de S\~ao Paulo (FAPESP),
the Israel Science Foundation founded by the Israel Academy of Sciences and 
Humanities, Istituto Nazionale di Fisica Nucleare (INFN),
the International Science Foundation (ISF), the National Science Foundation (Phy \#9602178), NATO (grant CR6.941058-1360/94),
the Russian Academy of Science,
the Russian Ministry of Science and Technology,
the Turkish Scientific and Technological Research Board (T\"{U}B\.ITAK),
the U.S. Department of Energy (DOE grant DE-FG02-91ER40664 and DOE contract
number DE-AC02-76CHO3000), and
the U.S.-Israel Binational Science Foundation (BSF).

\end{document}